# Long-Term Proportional Fair QoS Profile Follower Sub-carrier Allocation Algorithm in Dynamic OFDMA Systems

Arijit Ukil, Jaydip Sen, Debasish Bera
Wireless and Sensor Technology Innovation Lab, Tata Consultancy Services
BIPL, Sector-5, Saltlake, Kolkata- 700091, India
arijit.ukil@tcs.com

*Abstract*-- In this paper, Long-Term Proportional Fair (LTPF) resource allocation algorithm in dynamic OFDMA system is presented, which provides long-term QoS guarantee (mainly throughput requirement satisfaction) to individual user and follows every user's QoS profile at long-term by incremental optimization of proportional fairness and overall system rate maximization. The LTPF algorithm dynamically allocates the OFDMA sub-carriers to the users in such a way that in long-term the individual QoS requirement is achieved as well as fairness among the users is maintained even in a heterogeneous traffic condition. Here more than maintaining individual user's instantaneous QoS; emphasis is given to follow mean QoS profile of all the users in long-term to retain the objectives of both proportional fairness and multi-user raw rate maximization. Compared to the algorithms, which provide proportional fair optimization and raw-rate maximization independently, this algorithm attempts to provide both kinds of optimizations simultaneously and reach an optimum point when computed in long-term by exploiting the time diversity gain of mobile wireless environment.

## I. INTRODUCTION

Next generation broadband wireless applications require high data rate, low latency, minimum delay; in short highly demanding QoS. The capacity of a communication system is limited by its available resources like bandwidth and power and for a fixed bandwidth and power; the system capacity also becomes fixed. In a multi-user scenario, system performance optimization can not be guaranteed by optimizing only the individual link performance. Dynamic Resource Allocation is a kind of cross layer optimization mainly involving Physical (PHY) and Media Access Control (MAC) manages the system resources, like bandwidth, transmit power by exploiting the frequency and temporal dimension of the resource space adaptively to achieve the system performance objective.

Orthogonal Frequency Division Multiple Access (OFDMA) is the de facto standard multiple access scheme for next generation wireless standards like WiMAX, LTE, IMT-A. OFDMA, also referred to as Multi-user OFDM is normally characterized by a fixed number of orthogonal sub-carriers to be allocated to the available users [1, 2]. Sub-carrier allocation algorithms intelligently assign mutually disjoint sub-carriers to the users from apriori knowledge of the channel condition by taking the advantage of Multi User Diversity [3, 4, 5].

The optimization objective of the system is either to attain maximum aggregate capacity or to provide fairness among the users or to have a trade-off between the two optimization schemes. Raw-rate maximization algorithm maximizes the total system throughput but does not concern about starving users. In [6], the problem of dynamic subcarrier and power allocation with the objective of maximizing the minimum of the users' data rates subject to a total transmission power constraint is investigated. The fairness issue is emphasized in [7] by the concept of opportunistic scheduling. Tradeoff between capacity and fairness can be efficiently realized by efficient proportional fair (PF) OFDMA subcarrier and power allocation algorithms with QoS provisioning in [8].

From [6-8], it can be observed that the objectives of fair optimization and overall system throughput maximization are opposite in nature. It is also very much difficult and computationally expensive to satisfy each user's instantaneous data rate requirement. User satisfaction and service provider objective both can be preserved if individual user's QoS is maintained. As most of the current and next generation wireless applications do not degrade much if mean QoS guarantee is provided, where the mean is computed over few frame durations. From this perspective Long Term Proportional Fair (LTPF) algorithm is proposed which instead of achieving instantaneous fairness or capacity maximization, attempts to allocate the OFDMA subcarriers to the users to achieve its minimum mean data-rate requirement within a few frame duration or at least to follow the overall QoS profile by exploiting the time diversity gain. It produces better result in most cases than traditional PF optimization. QoS guarantee to the user is the most priority objective for next generation wireless broadband system. So LTPF algorithm is very much suitable and practical. Simulation results also justify the claim and show how QoS profile is closely followed in long term by the mean data rate achieved by LTPF algorithm.

The paper is organized as follows. The next section describes the system model. In section III instantaneous proportional fair optimization problem is discussed. LTPF optimization and algorithm are presented in detail in section IV. Simulation results and analysis of the LTPF algorithm are presented in the next section. Section VI provides the summary and conclusion.

## II. SYSTEM MODEL

Multiuser OFDMA system architecture with sub-carrier allocation module is shown in Fig. 1. Single cell wireless cellular network with one base station (BS) serving total K users is considered. The interference from adjacent cells is treated as background noise. The sub-carrier bandwidth is chosen to be sufficiently smaller than the coherence bandwidth of the channel in order to overcome frequency selective fading. Sub-carrier allocation is performed within the frame duration, which is assumed to be less than the channel coherence time. Perfect channel characteristic is assumed in the form of channel state information (CSI). LTPF algorithm is simplified by equally distributing the total available transmitted power as performance can hardly be deteriorated by equal power allocation to each subcarrier [5]. LTPF optimization is performed in long term duration, which is named as allocation duration. The mutually disjoint sub-carriers are denoted as: $\Omega_1, \Omega_2, ....\Omega_N$, where $\Omega_n$ = B/N and $\Omega_n \leq B_c$, where $B_c$ is the coherence Bandwidth of the channel and B is the total available bandwidth. $P_T$ is the total available transmit power and $P_{kn}$ is the transmit power for $n^{th}$ subcarrier when transmitted to $k^{th}$ user, where $P_{kn}$ = $P_T$/N . Total noise power density including background noise and AWGN noise is taken as $N_t$. Channel Gain for subcarrier n for user k at $t^{th}$ allocation instant is taken as $h_{knt}$. Minimum individual rate requirement for all the K users as per individual QoS is $[\gamma_1, \gamma_2, ....\gamma_K]$. Long-term allocation duration $\Delta_{ad}$ consists of a number of allocation epochs (=frame-duration) and $\Delta_{ad} = M \times T_f$, M = 1, 2 …, where $T_f$ is the frame-duration, which is taken as the unit allocation duration. $T_f \leq \Gamma_c$, where $\Gamma_c$ is the coherence time of the channel and $\Delta_{ad}$ should be (much) greater than $\Gamma_c$, in order to take the advantage of time diversity gain. Let $\omega_{kt}$ be the achievable rate for $k^{th}$ user at $t^{th}$ instant, then $\omega_{kt}$ is a function of the channel condition or $h_{knt}$. More specifically, $\omega_{kt}$ also depends on the sub-carrier allocation scheme considered and can be expressed as:

$$\omega_{kt} = \sum_{n=1}^{N} \Omega_n \times \rho_{knt} \times f(h_{knt}) \quad (1)$$

where $f(h_{knt})$ can be expressed as per (2) and $\rho_{knt}$ is the sub-carrier assignment matrix, which is equal to 1 if $n^{th}$ subcarrier assigned to $k^{th}$ user at $t^{th}$ time instant, else equal to 0 and $\sum_{k=1}^{K} \rho_{knt} = 1$.

$$f(h_{knt}) = \log_2\left(1 + \frac{h_{knt}^2 \times P_{kn}}{N_t \times \Omega_n}\right) \times SNR\_gap \quad (2)$$

SNR_gap is the imperfection of theoretical value of achievable data rate to the actual data rate and can be expressed as [10]: $\frac{-\ln(5 \times BER)}{1.6}$. Proportional Fairness Index (PFI) is a kind of optimized measure of user fairness. PFI at $t^{th}$ allocation instant for $k^{th}$ user is expressed as $pfi_{kt} = \frac{\omega_{kt}}{\overline{\omega_k}|_t}$. Total achievable data rate for $k^{th}$ user at the end of allocation duration $\Delta_{ad}$ is equal to $\omega_k$, which is equal to $\sum_{t=0}^{\Delta_{ad}} \omega_{kt}$.

## III. INSTANTANEOUS PROPORTIONAL FAIR OPTIMIZATION PROBLEM

PF in OFDMA system maximizes the sum of logarithmic mean user rates [8, 11].

$$pfi|_t = \max \sum_{n=1}^{N} \sum_{k=1}^{K} \ln \omega_{kt} \quad (3)$$

Equation (3) can be more generalized when PF is defined as follows [9]:

$$pfi|_t = \prod_{k \in K}\left(1 + \frac{\sum_{n \in N} \omega_{knt}}{(\Delta\tau - 1)\overline{\omega_{kt}}}\right) \quad (4)$$

where $pfi|_t$ is the proportional fairness index at $t^{th}$ instant, $\Delta\tau$ is average window size and $\overline{\omega_{kt}}$ is the average data rate achieved by the user k at the preceding allocation instant. Equation (4) is the optimal PF subcarrier allocation, can not be implemented in real systems due to its high computational complexity. The suboptimal form described in [11], where subcarrier n is allocated to k* user when:

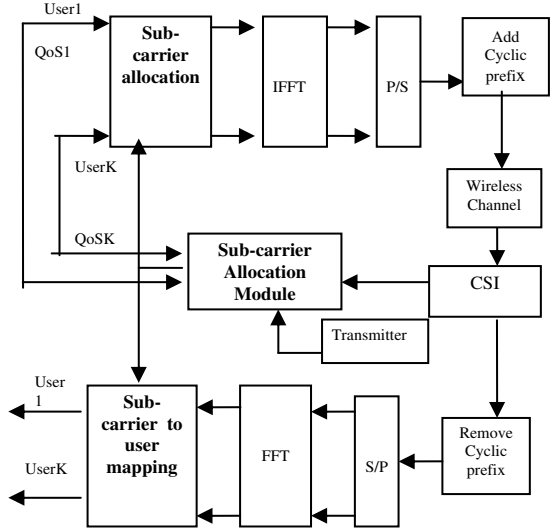

Fig. 1 Multiuser OFDMA System with Sub-carrier Allocation Module

$$k^* = \arg\max_k \frac{\omega_{kt}}{\overline{\omega_{kt}}} \quad (5)$$

$$\overline{\omega_{kt}} = (1 - \frac{1}{\Delta\tau})\overline{\omega_{k(t-1)}} + \frac{1}{\Delta\tau}\omega_{k(t-1)} \quad (6)$$

System constraint to provide QoS is:

$$\omega_{kt} \geq \gamma_k, \quad \forall k \quad (7)$$

The optimization and sub-optimization schemes (3-7) discussed so far deals with maintaining instantaneous QoS guarantee, which according to the proposed LTPF optimization does not offer the best system or user level performance gain in wireless mobile environment.

## IV. LONG TERM PROPORTIONAL FAIR OPTIMIZATION AND ALGORITHM

According to equation (3-7) PF metric is calculated from instantaneous achievable rate and sub-carrier allocation decision taken at every allocation epoch ($T_f$) depends on the instantaneous channel condition $h_{knt}|_{t=T_f}$ and the fairness is maintained according to the law of dimishing return (3). The channel dynamics of mobile wireless broadband system is very high. So instantaneous decision making may not be the best way for an optimization scheme. Instead LTPF algorithm proposes long term maximization of user's mean achievable data rate subject to minimum data rate constraint. LTPF incurs the advantage of time diversity (TD) gain, when QoS parameter computation time ($\Delta_{ad}$) is more than the channel coherence time. But TD technique is restricted to delay-tolerant applications in mobile wireless environment. Let Probability of error of detecting x is given by:

$$P_e(x) = Q(\sqrt{2xSNR})$$

where Q(.) is complementary cumulative distribution function of a $N(0, N_T)$ random variable and $\overline{P_e}$ be the average probability of error of received signal, it can be proved that [13]:

$$\overline{P_e} \approx \binom{2M-1}{M} \times \frac{1}{4(SNR)^M} \quad (8)$$

where M is the diversity branch or in this case, the number of unit allocation instants taken in PFI computation.

Equation (8) shows that substantial performance gain can be achieved by taking advantage of time diversity gain. LTPF optimization utilizes this performance gain for its attempt to converge to $\gamma_k$ by computing PFI over $\Delta_{ad}$ as well as relaxing the QoS constraint of minimum data rate as a mean value. Basically, instead of instantaneous optimization, LTPF allocates OFDMA sub-carriers to optimize the performance over few allocations and assures QoS guarantee in an average basis within that pre-defined allocation duration. The idea is formulated below:

$$\max \sum_{t=0}^{\Delta_{ad}} \sum_{n=1}^{N} \sum_{k=1}^{K} \frac{\omega_{knt}}{\overline{\omega_k}\big|_{t=0}^{\Delta_{ad}}} \quad (9)$$

$$\omega_k = \sum_{t=0}^{\Delta_{ad}} \sum_{n=1}^{N} \Omega_n \times \rho_{knt} \times f(h_{knt}) \quad (10)$$

$$\sum_{t=0}^{\Delta_{ad}} \sum_{n=1}^{N} \sum_{k=1}^{K} \rho_{knt} = N \times M \quad (11)$$

subject to:

$$\overline{\omega_k} \geq \gamma_k, \quad \forall k \quad (12)$$

Equations (9-12) describes LTPF optimization scheme. It can be noticed that more the value of M, the more ergodic the optimization scheme becomes; at M=1, the optimization is purely proportional fair. It can also be noted from (8) that more time diversity gain would be achieved with the increasing value of M, i.e.

$$\lim_{\Delta_{ad} \to \infty} P(\overline{\omega_k} \to \gamma_k) = 1 \quad (13)$$

Equation (13) is an important result of LTPF, which states that with more allocation instants considered, more is the probability of mean achievable data rate to converge to the QoS requirement. This is due to the fact that in wireless mobile environment over long duration, time diversity gain becomes high and the mean channel condition ($\overline{h_{kn}}$) follows similar distribution according to the Bernoulli's Law of Large Numbers, which justifies the intuitive interpretation that the expected value of a random variable is basically long-term average when sampled repeatedly. As $h_{kn}$ is commonly considered as idenpendepent and identically distributed (i.i.d) random variable with mean $\mu_k$, then theoretically,

$$\lim_{\Delta_{ad} \to \infty} TDgain = \mu_k \quad (14)$$

There exists some limitations or constraints for (13) to be true. Let the long term average data rate of $k^{th}$ user equals to $\overline{\omega_k} = \lim_{\Delta_{ad} \to \infty} f(\mu_k)$, then the following condition (15) should hold true for (13) to be valid.

$$|\gamma_k - \overline{\omega_k}| \leq \varepsilon \quad (15)$$

where $\varepsilon$ is a very small number.

In practical systems, QoS guarantee has to be made within few frame durations. This limits the value of $\Delta_{ad}$, consequently the probability of convergence towards $\gamma_k$. The converging rate towards $\gamma_k$ increases at large $\Delta_{ad}$ as well as in highly mobile environment. So the value of $\Delta_{ad}$ should be chosen as large as practically possible. Based on the LTPF optimization as per (9-12) the proposed LTPF algorithm is described as below:

1. Initialize $\overline{\omega_k}\big|_{t=0} = \psi$, $pfi_{knt}\big|_{t=0} = 0$, $\omega_{kt} = 0$, $\forall k$, where $\psi$ is a random number.

2. Assign initial subcarriers to the users from proportional fairness index, at $t = 0$.
   while ($\{\Omega_n\} \neq \Phi$) do
     for k= 1: K
       calculate $\omega_{knt}$
       $$pfi_{knt} = \frac{\omega_{knt}}{\overline{\omega_k}\big|_{t=0}^{T_f}}$$
     end for
     $k^* = \arg\max_k (pfi_{knt})$
     $\omega_{k^*t} = \omega_{k^*nt} + \omega_{k^*t}$
     $\overline{\omega_k}\big|_{t=0}^{T_f} = \omega_{kt}$
   end while

3. Incorporate long term notion and time diversity gain
   for $t = T_f : \Delta_{ad}$
     while ($\{\Omega_n\} \neq \Phi$) do
       for k= 1: K
         if ($\overline{\omega_k} < \gamma_k$)
           calculate $\omega_{knt}$
           $$pfi_{knt} = \frac{\omega_{knt}}{\overline{\omega_k}\big|_{t=T_f}^{t}}$$
         else
           $pfi_{knt} = -\infty$
         end for
         $k^* = \arg\max_k (pfi_{knt})$
         $\omega_{k^*t} = \omega_{k^*nt} + \omega_{k^*t}$
       end while
       calculate $\overline{\omega_k}\big|_{t=T_f}^{t} = \overline{\omega_k}\big|^{t} - \overline{\omega_k}\big|_{t=0}^{T_f}$
   end for

V. SIMULATION RESULT AND ANALYSIS

In this section we present simulation results of the proposed LTPF algorithm under the system parameters and simulation scenario given in Table 1. The system parameters are roughly based on Mobile WiMAX Scalable OFDMA-PHY. Achievable data rate of individual user is calculated according to (2). Frequency reuse factor = 1 is taken, so that all the available sub-carriers can be allocated to the users.

TABLE I
Simulation and System Parameters

| Available Bandwidth | 1.25 MHz |
|---|---|
| Total Transmitted Power | 20 dBm |
| Number of users | 20 |
| Number of sub-carriers | 72 |
| BER | $10^{-3}$ |
| Frame duration | 5 msec |
| Allocation instant (Tf) | Frame duration |
| $\Delta_{ad}$ | $M \times T_f$ |
| Channel model | Rayleigh |
| Modulation | 16QAM |
| Channel sampling frequency | 1.5 MHz |
| Maximum Doppler | 100Hz |

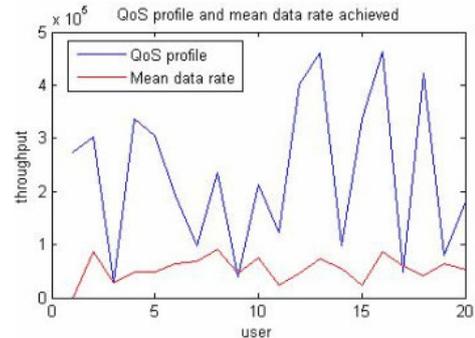

Figure 2: QoS profile and mean data rate achieved when M=1

The simulation results depict the performance of the algorithm and also show how user data rate follows the overall QoS profile at long term. By overall QoS profile, we mean the plot of every user's fixed average data-rate requirement within the long term duration ($\Delta_{ad}$). Heterogeneous traffic model with variable QoS demand by the user is considered, as in a broadband wireless scenario it is very much practical to assume large differences in QoS requirement.

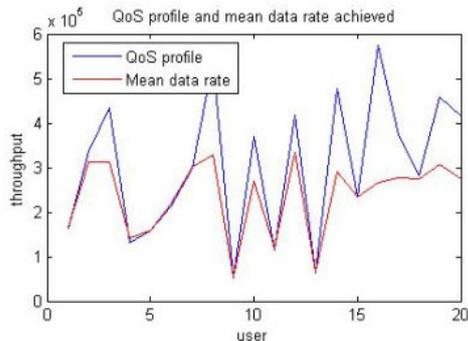

Figure 3: QoS profile and mean data rate achieved when M =4

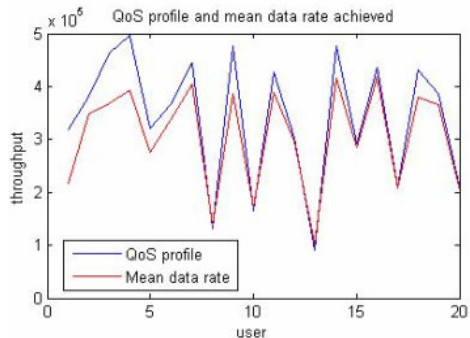

Figure 4: QoS profile and mean data rate achieved when M =10

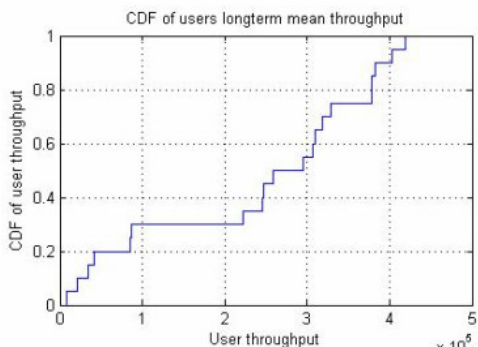

Figure 5: CDF plot of the user mean data rate at M =10

Fig. 2 depicts the plot comparing QoS profile of users' achieved mean data-rate when $\Delta_{ad}$ equals to one unit. Here it is clear that the achieved mean data-rate profile deviates from the QoS profile considerably and also the individual data-rate achieved is substantially low, as in this case, LTPF algorithm becomes purely proportional fair. Fig. 3-4 show the result when LTPF optimization is incorporated in sub-carrier allocation. Fig. 3 and 4 clearly show that the mean achieved data-rate is considerably improved and attempts to follow the QoS profile. Fig. 4 establishes the fact that achievable mean data rate is converging towards $\gamma_k$, when $\Delta_{ad}$ is of large magnitude. Fig. 5 is the corresponding CDF (Cumulative Distribution Function) plot.

## VI. SUMMARY AND CONCLUSION

The LTPF algorithm has shown the characteristics of long term QoS profile follower by taking the advantage of time diversity gain. This algorithm particularly performs better optimization in mobile environment and delay-tolerant applications. Simulation results depict the performance of the algorithm. The distinguished quality of LTPF as QoS profile follower is of very much practical importance and can well be implemented for next generation broadband wireless systems like LTE, WiMAX, IMT-A for acheiving the target performance.